\documentclass[a4paper,fleqn,usenatbib]{mnras}

\usepackage[T1]{fontenc}
\usepackage{ae,aecompl}

\usepackage{graphicx}	
\usepackage{amsmath}	
\usepackage{amssymb}	

\title[X-ray flare after GRB 170817A]{Could an X-ray Flare after GRB 170817A Originate from a Post-merger Slim Accretion Disc?}

\author[Lin et al.]{
Yi-Qing Lin,$^{1}$
Zi-Gao Dai,$^{2,3}$
Wei-Min Gu$^{4}$
\\
$^{1}$Fujian Provincial Key Laboratory of Optoelectronic Technology
and Devices, Xiamen University of Technology,\\ Xiamen, Fujian 361024, China;
yqlin@xmut.edu.cn\\
$^{2}$School of Astronomy and Space Science, Nanjing University,
Nanjing 210046, China; dzg@nju.edu.cn\\
$^{3}$Key Laboratory of Modern Astronomy and Astrophysics (Nanjing University), Ministry of Education, China\\
$^{4}$Department of Astronomy, Xiamen University, Xiamen,
Fujian 361005, China; guwm@xmu.edu.cn
}

\date{Accepted XXX. Received YYY; in original form ZZZ}

\pubyear{2019}

\begin{document}
\label{firstpage}
\pagerange{\pageref{firstpage}--\pageref{lastpage}}
\maketitle

\begin{abstract}
GRB 170817A, detected by Fermi-GBM 1.7\,s after the merger of a neutron star (NS) binary, provides the first direct evidence for a link between such a merger and a short-duration gamma-ray burst. The X-ray observations after GRB 170817A indicate a possible X-ray flare with a peak luminosity
$L_{\rm peak} \sim 2\times 10^{39}\,{\rm erg\,s}^{-1}$ near day 156. Here we show that this X-ray flare may be understood based on a slim disc around a compact object. On the one hand, there exists the maximal accretion rate $\dot M_{\rm max}$ for the slim disc, above which an optically thick outflow is significant and radiation from the disc is obscured. Based on the energy balance analysis, we find that $\dot M_{\rm max}$ is in the range of $\sim 4\dot M_{\rm Edd}$ and $\sim 21\dot M_{\rm Edd}$ when the angular velocity of the slim disc is between $\rm (1/5)^{1/2}\Omega_K$ and $\rm \Omega_K$ (where $\dot M_{\rm Edd}$ is the Eddington accretion rate and $\Omega_K$ is the Keplerian angular velocity). With $\dot M_{\rm max}$, the slim disc can provide a luminosity $\sim L_{\rm peak}$ for a compact object of $2.5 M_{\sun}$. On the other hand, if the merger of two NSs forms a typical neutrino-dominated accretion disc  whose accretion rate $\dot M$ follows a power-law decline with an index $-1.8$ , then the system must pass through the outflow regime and enter the slim disc in $\sim 11-355$ days. These results imply that a post-merger slim accretion disc could account for the observed late-time $L_{\rm peak}$.
\end{abstract}




\section{Introduction}

Gamma-ray bursts (GRBs) are short-duration flashes of gamma-rays
 occurring at cosmological distances.
An extremely low-luminosity GRB 170817A was detected by Gamma-ray Burst Monitor (GBM) on
 board the Fermi satellite at 12:41:04.446 UTC as a short
 GRB \citep{Goldstein2017}.
 This burst is highly noticeable due to its connection to
 the gravitational wave event GW170817 which was detected by the Laser Interferometer Gravitational-wave
  Observatory (LIGO) approximately 1.7\,s before the GBM triggered \citep{Abbott2017}.
  A few hours later, GW170817 was further identified to originate from
  a binary neutron star (NS) merger, thanks to the discovery of an associated kilonova \citep{Arcavi2017}.
   This is the first direct evidence for a link between
NS-NS mergers and short GRBs.
In addition, the gamma-ray signal of GRB 170817A and the following
emission are unlikely to be those of any other short GRB seen before \citep{Kasliwal2017, Matsumoto2019}.
  Compactness arguments reveal that the observed gamma-rays
  may be produced in a mildly or fully relativistic outflow \citep{Kasliwal2017}.
There are also arguments that the gamma-rays may be generated when a cocoon
breaks out from the merger ejecta,
where the cocoon arises from either an emerging or a choked jet
  \citep{Gottlieb2018, Nakar2018}.
 A plausible explanation for the observed emission is synchrotron radiation
  from the fast tail of dynamical ejecta during the merger
  or from electrons accelerated through a forward shock driven
  by the merger outflow \citep{Hotokezaka2018, NakarPiran2018}.

 X-ray flares are erratic temporal features,
 commonly seen in GRB afterglows,
 and have been observed both in long and short GRBs
 \citep[e.g.,][]{Romano2006, Campana2006, Falcone2006, Margutti2011}.
 They usually happen at $10^2-10^5\rm s$ after the prompt emission,
 a few flares can occur even up to several days after the
 GRB trigger \citep[e.g.,][]{Chincarini2007, Chincarini2010, Falcone2007}.
The fluence of an X-ray flare is usually smaller than that of the prompt emission
and the temporal behaviour of flares is quite similar to the prompt
 emission pulses. Thus, X-ray flares may have the same physical
 origin as the prompt pulses \citep{Burrows2005, Falcone2006,
 Falcone2007, Zhang2006, Nousek2006, Liang2006, Chincarini2007, Chincarini2010,
 Wu2013, Yi2015}, and may provide
 an important clue to understand the mechanism of GRB phenomenon.
The physical origin of X-ray flares
remains mysterious, including
late-time activity of the central
 engines \citep[e.g.,][]{Falcone2007, Chincarini2007, Chincarini2010, Bernardini2011, Mu2016, Mu2018},
fragmentation of the accretion disc \citep{Perna2006},
intermittent accretion behaviour caused by a time variable
magnetic barrier \citep{Proga2006},
magnetic reconnection from a post-merger millisecond
pulsar \citep{Dai2006}
and magnetic dissipation in a decelerating shell \citep{Giannios2006}.

The X-ray observations have shown that GRB 170817A has an X-ray flare occurring
 between day 155 and 157 after prompt emission, followed by a rapid decay phase,
 which suggests that the X-ray emission
 peaked at day 156 \citep{Piro2019}.
  The peak time $ t_{\rm peak}\approx \rm 156\,d$
  and the peak luminosity $L_{\rm peak}\sim 2\times 10^{39}\rm erg\;s^{-1}$
  with the distance $D_L=40\,\rm Mpc$,
   fall within the expected range of values derived by an extrapolation
    the flux distribution of GRB X-ray flares to later times
     \citep{Bernardini2011,Piro2019}.
The idea of X-ray emission from late-time accretion around a neutron
       star merger remnant have been addressed extensively
       generally \citep{Rosswog2007, Metzger2010,Fernandez2015}
      and in the context of GRB 170817A \citep{Kisaka2015, Matsumoto2018}.
      Here we show that the late-time X-ray flare
      after GRB 170817A may be understood based on a slim disc model.

The slim disc model \citep{Abramowicz1988}
is introduced for super-Eddington accretion flows
with the mass accretion rate $\dot M > \dot M_{\rm Edd}$,
where $\dot M_{\rm Edd}=64\pi GM/{c\kappa _{\rm es}}$.
Here $M$ is the mass of a black hole (BH), $c$ is the speed of light,
and $\kappa _{\rm es}=0.4\,\rm cm^2\,g^{-1}$ is the
electron scattering opacity. The
local analysis reveals that there may exist the maximal accretion rate
$\dot M_{\rm max}(r)$ at a radius of the slim disc \citep{Gu2007,Cao2015}.
Global transonic solutions for slim discs
with the explicit vertical gravitational force confirm
that the possible maximal accretion rate indeed occurs for a slim disc
\citep{Jiao2009}.
The physical reason for this may be related to
the limited cooling by advection and radiation and therefore
no thermal equilibrium can be established for $\dot M > \dot M_{\rm max}$
\citep{Gu2015}.
As a consequence, outflows ought to occur in such flows.
The previous simulations have also found strong outflows for extremely
high accretion rates \citep[e.g.,][]{Ohsuga2011,Sadowski2016}.

In this paper, we focus on the timescale and the luminosity of
the X-ray flare after GRB 170817A and show that such properties may be understood
based on a slim accretion disc around a post-merger compact object.
This paper is organized as follows.
The physical model is presented in $\S 2$.
Results are presented in $\S 3$.
Conclusions and discussion are given in $\S 4$.

\section{Physical model}

If the merger of double NSs forms a typical neutrino-dominated
accretion disc, then the accretion rate $\dot M$ maybe follows a power-law decline
with an index of $-5/3$ \citep {Rees1988}.
Recently, a power-law fit to the accretion
rate in the GRMHD (general-relativistic magnetohydrodynamic)
model for $t > 1\rm s$ yields $t^{-1.8}$
\citep {Fernandez2019, Siegel2018}, that is,
\begin {equation}
 \dot M= \dot M_\nu(t/t_\nu)^{-1.8},
\end{equation}
and thus the timescale towards the slim disc regime $t_{\rm ph}$ can be
estimated as
\begin {equation}
t_{\rm ph}=t_\nu( \dot M_\nu /\dot M_{\rm ph})^{5/9},
\end{equation}
where $t_\nu=2\,\rm s$ is the duration of GRB 170817A, and
 $\dot M_\nu$ is the accretion rate of the neutrino-dominated accretion disc
 in the range of $0.001-0.1M_{\sun} \rm s^{-1}$.
Here, the lower limit of $0.001M_{\sun} {\rm s^{-1}}$ is known as
the ignition accretion rate \citep{Chen2007}, and the upper limit of
$0.1M_{\sun} {\rm s^{-1}}$ can be estimated by the prompt gamma-ray
emission duration $2\,{\rm s}$ and an upper limit of the disc mass
$\sim0.2 M_{\sun}$.
Thus, if the typical accretion rate $\dot M_{\rm ph}$ is given,
we can predict the time of reaching the slim disc regime.
We derive such a typical rate $\dot M_{\rm ph}$ as follows.

In the energy balance equation for super-Eddington accretion flows,
the viscous heating rate equals the cooling rate, where
the cooling mechanisms include advection, radiation and outflows, and therefore we have
\begin {equation}
Q_{\rm vis}^+ =Q_{\rm rad}^-+Q_{\rm adv}^- + Q_{\rm outflow}^-,
\end{equation}
where $Q_{\rm vis}^+$ is the viscous heating rate,
$Q_{\rm rad}^-$ is the radiative cooling rate,
$Q_{\rm adv}^-$ is the advective cooling rate,
and $Q_{\rm outflow}^-$ is the cooling rate by outflows.

For super-Eddington accretion flows with high
accretion rates, the advective cooling rate may be limited to be
under $30\%$ of the viscous heating rate \citep{Gu2015}.
We assume that a necessary condition for the slim
disc regime is that the radiative cooling should be significant.
Otherwise, the cooling ought to be dominated by outflows, in which case
the radiation of the disc is obscured by the
dominant optically-thick outflows.
Thus, we assume the following condition,
\begin {equation}
Q_{\rm rad}^- \ga \frac {1}{2}Q_{\rm vis}^+.
\end{equation}

In general, we have a relation between the radiative cooling
rate and the vertical energy flux,
\begin {equation}
Q_{\rm rad}^-=2F_{\rm z},
\end{equation}
where $F_{\rm z}$ is the radiative energy flux in $z$-direction.
The gravitational force in the vertical direction, $\kappa_{\rm es}/{c}\cdot F_{\rm z}$, has a maximum value
\citep{Cao2015}
\begin {equation}
\frac {\kappa_{\rm es}}{c}\cdot F_z \la \left (\frac{d\Phi} {dz}
\right)_{\rm max},
\end{equation}
where $\Phi$ is the gravitational potential. The vertical component of gravity
of the BH increases from the mid-plane of the disc along $z$,
and reaches the maximum at $z/R=\sqrt 2/2$, where $R$ is the radius \citep{Cao2015}.

We adopt the Paczy\'nski-Wiita potential \citep{Paczynski1980},
\begin {equation}
\Phi=-\frac{\rm GM}{\sqrt{(R^2+z^2)}-R_{\rm g}},
\end{equation}
where $R_{\rm g} \equiv 2GM/c^2$ is the Schwarzschild radius.
The expression for $Q_{\rm vis}^+$ is
\begin {equation}
Q_{\rm vis}^+ =\frac{3} {8 \pi} \dot M \Omega^2 \left (1-\frac{j}
{\Omega_{\rm K} R^2}\right )\left (-\frac {2 d\ln\Omega_{\rm K}}{3 d\ln R}
\right),
\end{equation}
where $j$ is an integration constant determined by the zero-torque boundary
condition at the last stable orbit, representing the specific
angular momentum per unit mass of the material accreted by the BH,
$\Omega$ is the angular velocity, and $\Omega_{\rm K}^2=GM/[R(R-R_{\rm g})^2]$.
It is known that the slim disc rotates in a
sub-Keplerian manner \citep{Abramowicz1988}.
We therefore assume that $\Omega=\lambda \Omega_{\rm K}$ and $0 < \lambda < 1$,
where the parameter $\lambda$ can be estimated from previous self-similar
solutions.

\section{Results} \label{sec:results}

\begin{figure}
\includegraphics[angle=0,width=8cm]{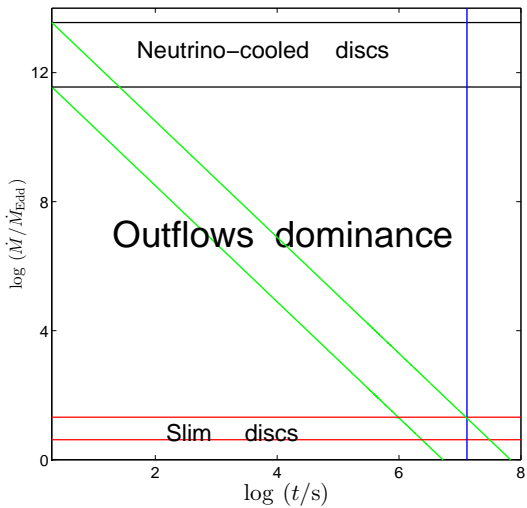}
\caption{
Accretion rate $\dot M$ as a function of time $t$.
From top to bottom, four horizontal lines represent the accretion rate
as $0.1M_{\sun} {\rm s^{-1}}$, $0.001M_{\sun} {\rm s^{-1}}$,
$21\dot M_{\rm Edd}$ and $4\dot M_{\rm Edd}$, respectively.
The vertical blue line corresponds to 156 days after the GRB.
The two green lines represent a power-law decline of the accretion rate
with an index of $-1.8$.}
\label{f1}
\end{figure}

Based on Equations~(3)-(8),
we take $R=10R_{\rm g}$ as a typical radius for analysis,
$\lambda=(1/5)^{1/2}$ for slim discs \citep{Wang1999} ,
and $j$ is assumed to be $1.8cR_{\rm g}$.
As a consequence, the maximal accretion rate can be
derived as $\dot M_{\rm max} = 21\dot M_{\rm Edd}$,
for which Inequality~(4) matches the boundary $Q_{\rm rad}^- = Q_{\rm vis}^+ /2$.
If the slim disc rotates in a Keplerian manner,
then the maximal accretion rate is $4\dot M_{\rm Edd}$.
Figure~1 shows the accretion rate $\dot M$
as a function of time $t$.
The two red horizontal lines represent the maximal accretion rate
$21\dot M_{\rm Edd}$ and $4\dot M_{\rm Edd}$, and the blue vertical line
corresponds to 156 days.
For the accretion rates, there are two regions for thermal equilibrium
solutions without strong outflows.
One region has a relatively high accretion rate, $0.001-0.1 M_{\sun} \rm s^{-1}$
 (two black horizontal lines),
corresponding to the neutrino-cooled discs for the central engine of GRBs.
The other has a relatively low rate, $\dot M_{\rm Edd} \la \dot M \la \dot M_{\rm max}$,
corresponding to the photon radiation-dominated discs, where X-ray radiation
is the dominant mechanism. The two regions are plotted in Figure 1.
As shown by Figure~1, between these two regions,
outflows ought to be optically thick and dominance, so that the
radiation from the disc is obscured.
The two green lines show that the accretion rates
follow a power-law decline with an index of $-1.8$.
The cross points of the $\rm 21\dot M_{Edd}$ and
the two green lines are $t \simeq 11$ and 145 days.
The cross points of the $\rm 4\dot M_{Edd}$ and
the two green lines are $t\simeq 28$ and $355$ days.
It is seen from Figure~1 that
the cross point of the red lines and the blue line (156 days) is nearly located
between the two green lines.
Our calculations above do not take into account
the spin of the BHs.
If the spin of the BHs is considered,
the energy release efficiency of accretion discs should be improved,
so that the maximum accretion rate will be reduced,
which makes the intersection time longer,
which means that the time of the accretion disc
reaching the slim disc, according to our model, is in agreement with
the observations.

Therefore, our scenario for the observed late-time X-ray flare is described as follows.
After the prompt gamma-ray emission, the accretion rate varies following
a power-law decline with an index of $-1.8$, and thus the accretion
flow will first enter an outflow-dominated regime
($21 \dot M_{\rm Edd} - 0.001 M_{\sun}~{\rm s}^{-1}$ or
$4 \dot M_{\rm Edd} - 0.001 M_{\sun}~{\rm s}^{-1}$).
In such a case,
the dominant optically thick outflow can prevent the X-ray emission
of the accretion flow from being observed. That is why the X-ray luminosity
is quite low for the system in the ``outflow-dominated" region (Figure~1),
even though the accretion rate may be extremely high.
On the contrary, when the accretion approaches the maximal rate
of a slim disc, the outflow will become weak and the observed X-ray emission
from the accretion flow will reach the peak luminosity. Finally,
following the decline in the accretion rate, the inflow will become around
Eddington and then sub-Eddington. Consequently, the X-ray luminosity will
drop to a low level. From the physical point, the observed X-ray flare
is related to a power-law decline for the accretion rate and the
occurrence of a strong optically thick outflow for an accretion rate
beyond the maximal value of a slim disc.

\begin{figure}
\includegraphics[angle=0,width=8cm]{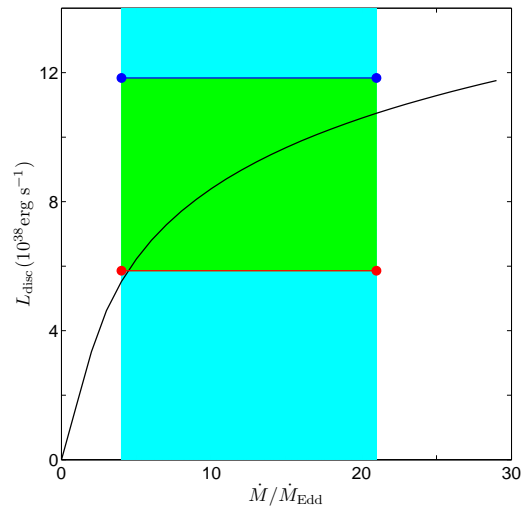}
\caption{
Disc luminosity $L$ as a function of accretion rate
$\dot M$. The black line represents the analytic relation
of Equation~(9).The light blue shaded region
shows the maximum accretion rate between
$4\dot M_{\rm Edd}$ and $21\dot M_{\rm Edd}$.
The two horizonal lines represent
the luminosity $L_{\rm disc}\sim 1.2\times 10^{39}\rm erg\; s^{-1}$ (blue line) and
  $5.8\times 10^{38} \rm erg\; s^{-1}$ (red line)
  corresponding to different half-thickness
at the maximal accretion rate from $4\dot M_{\rm Edd}$
to $21\dot M_{\rm Edd}$.
}
\label{f2}
\end{figure}

Now we investigate the typical luminosity when the accretion enters the slim disc region.
Some previous theoretical works have provided
the disc luminosity as a function of accretion rate
\citep[e.g., Equation~10.27 of][]{Kato2008},
\begin{equation}
 \frac{L}{L_{\rm Edd}} \simeq
\left\{
             \begin{array}{lr}
              1 + \ln (\frac{8}{15} \dot m ),\qquad \dot m \geqslant \frac{15}{8}, \\
             \frac{8}{15} \dot m, \qquad \dot m < \frac{15}{8}, \\

             \end{array}
\right.
\end{equation}
where $\dot m=\dot M/\dot M_{\rm Edd}$ and the difference in efficiency for the Eddington accretion rate ($1/16$)
is taken into consideration.
Such a relation is shown by the black solid line in Figure~2.

\citet{Narayan2016} showed that the isotropic
     equivalent luminosity $L_{\rm iso}$ of the supercritical
 BH accretion model is related to the inclination angle $\theta$.
       So, for the luminosity of the X-ray flare
       after GRB 170817A near day 156,
       we should consider the factor $ 1- \cos \theta $.
        Since $\sim 2\times 10^{39}\rm erg\;s^{-1}$ is around
 the typical luminosity provided by a slim disc around a compact object
  (e.g., stellar-mass BH or NS).
The half-thickness of the slim disc may range from
 $H/R\sim 1/\sqrt 5$ \citep{Wang1999} to
 $H/R\sim 1$ \citep{Abramowicz1995},
 and then the corresponding luminosity $L_{\rm disc}$ rangs from
  $\sim 1.2\times 10^{39}\rm erg\;s^{-1}$
  to $\sim 5.8\times 10^{38}\rm erg\;s^{-1}$ .
      The two horizonal lines represent
the luminosity $L_{\rm disc}$
  $1.2\times 10^{39} \rm erg\; s^{-1}$ (blue line) and
  $5.8\times 10^{38} \rm erg\; s^{-1}$ (red line)
at the maximal accretion rate from $4\dot M_{\rm Edd}$
to $21\dot M_{\rm Edd}$. The shaded part of light blue
shows the maximum accretion rate between
$4\dot M_{\rm Edd}$ and $21\dot M_{\rm Edd}$.
The intersection of the two horizonal lines and light blue
shadows are represented by the green shadow region.
We can see that the black solid line
lies in the middle of this green shadow region.

 This means that, if the vertical advection process
can play a role, then the radiation from the maximal accretion rate
can well explain the observed peak luminosity of the X-ray flare.
Thus, we suggest that the central engine for the observed late-time
X-ray flare may be related to the super-Eddington accretion process.

\section{Conclusions and discussion}

We have proposed that if the accretion rate follows a power-law
decline with a typical index of $-1.8$, then the observed late-time X-ray
flare after GRB 170817A around day 156 can be
well explained by a slim disc surrounded by a stellar-mass BH.
If the vertical advection process found
in simulations can play a role, a slim disc with the maximal accretion rate
can provide the observed peak luminosity. We therefore suggest
super-Eddington accretion as a central engine of the X-ray flare.
The physics of the observed late-time X-ray flare
is related to a power-law decline for the accretion rate and the occurrence
of a strong optically thick outflow for an accretion rate beyond
the maximal value of a slim disc.

On the other hand, if the central object is a neutron star rather than
a black hole, the slim disc model may still work except for the most inner
regions. Since the neutron star has a hard surface instead of a horizon,
the trapped photons in the slim discs cannot be absorbed by the neutron
star. A possible scenario is that, for the same super-Eddington accretion
rate, outflows in a neutron star system may be even stronger, and they
can carry away the trapped photons outwards and a part of the photons can
be released far away from the disc. Following this argument, the slim disc
around a neutron star may also be responsible for the late-time X-ray flare.
The difference may be that the maximal accretion rate
in the neutron star system may be lower than the black hole case, owing to
the stronger outflows in the former system.

It is generally believed that GRB 170817A originates from
a binary NS merger. However, strictly speaking, the possibility of
a light BH-NS merger isn't ruled out. The serious argument against
a BH-NS merger is that in this case the BH mass would be unusually low,
$\sim 1.4 M_{\sun}$. In fact, a NS-mass BH may be present in relativistic
binary systems because such binaries could result from collisions of primordial BHs with NSs
\citep{Abramowicz2018}. In the BH-NS merger case, compact objects after the mergers
are undoubtedly black holes, and therefore our slim disc model is also valid.

\section*{Acknowledgements}

The authors would like to thank the referee for constructive suggestions.
This work was supported by the National Key Research and Development Program of China (under grant No. 2017YFA0402600) and the National Natural Science Foundation of China (under grant No. 11573014, 11833003 and 11573023).





\bsp	
\label{lastpage}
\end{document}